\documentclass[
prb,
reprint,
superscriptaddress,
]{revtex4-2}

\usepackage{amsmath,amssymb} 
\usepackage{amsmath} 
\usepackage{amsfonts} 
\usepackage{bm} 
\usepackage[dvipdfm,colorlinks=true,citecolor=blue,linkcolor=blue,filecolor=blue,urlcolor=blue]{hyperref} 
\usepackage[dvipdfmx]{graphicx} 
\usepackage{braket} 
\usepackage{ulem}
\usepackage[usenames,dvipsnames]{xcolor}

\begin{document}

\title{Multipolar Phase Transition in the 4$f^2$ fcc Lattice Compound PrCdNi$_{4}$}

\author{Yuka Kusanose}
	\email{kusanose.yuka.p7@f.mail.nagoya-u.ac.jp}
	\affiliation{
	Department of Quantum Matter, Graduate School of Advanced Science and Engineering, Hiroshima University, Higashi-Hiroshima 739-8530, Japan}
	\affiliation{
	Department of Applied Physics, Nagoya University, Nagoya 464-8603, Japan}
\author{Yasuyuki Shimura}
	\affiliation{
	Department of Quantum Matter, 
	Graduate School of Advanced Science and Engineering, Hiroshima University, Higashi-Hiroshima 739-8530, Japan}
\author{Kazunori Umeo}
	\affiliation{Natural Science Center for Basic Research and Development (N-BARD),
	Hiroshima University, Higashi-Hiroshima 739-8526, Japan}
\author{Naomi Kawata}
	\affiliation{Natural Science Center for Basic Research and Development (N-BARD),
	Hiroshima University, Higashi-Hiroshima 739-8526, Japan}
\author{Toshiro Takabatake}
	\affiliation{
	Department of Quantum Matter, 
	Graduate School of Advanced Science and Engineering, Hiroshima University, Higashi-Hiroshima 739-8530, Japan}
\author{Taichi Terashima}
	\affiliation{Research Center for Materials Nanoarchitectonics (MANA), National Institute for Materials Science (NIMS), Tsukuba 305-0003, Japan}
\author{Naoki Kikugawa}
	\affiliation{Center for Basic Research on Materials (CBRM), National Institute for Materials Science (NIMS), Tsukuba 305-0003, Japan}
\author{Takako Konoike}
	\affiliation{Research Center for Materials Nanoarchitectonics (MANA), National Institute for Materials Science (NIMS), Tsukuba 305-0003, Japan}
\author{Yuya Hattori}
	\affiliation{Center for Basic Research on Materials (CBRM), National Institute for Materials Science (NIMS), Tsukuba 305-0003, Japan}
\author{Kazuhiro Nawa}
	\affiliation{
	Institute of Multidisciplinary Research for Advanced Materials, Tohoku University, Sendai 980-8577, Japan}
\author{Hung-Cheng Wu}
	\altaffiliation[Present address: ]{Department of Physics, National Sun Yat-sen University, Kaohsiung, 80424, Taiwan}
	\affiliation{
	Institute of Multidisciplinary Research for Advanced Materials, Tohoku University, Sendai 980-8577, Japan}
\author{Taku J Sato}
	\altaffiliation[Present address: ]{Neutron Science Laboratory, Institute for Solid State Physics, University of Tokyo, 5-1-5 Kashiwanoha, Chiba 277-8581, Japan}
	\affiliation{
	Institute of Multidisciplinary Research for Advanced Materials, Tohoku University, Sendai 980-8577, Japan}
\author{Takahiro Onimaru}
	\affiliation{
	Department of Quantum Matter, 
	Graduate School of Advanced Science and Engineering, Hiroshima University, Higashi-Hiroshima 739-8530, Japan}

\date{\today}


\begin{abstract}
Transport and magnetic properties of a 4$f^{2}$ fcc lattice compound, PrCdNi$_4$, were studied.
The magnetic susceptibility, $\chi(T)$, follows the Curie--Weiss law from 300 K to 20 K, as expected for a free Pr$^{3+}$ ion.
As the temperature decreases below 5 K, $\chi(T)$ approaches a constant, indicating van-Vleck paramagnetic behavior.
The specific heat, $C(T)$, displays a broad shoulder at around 4 K, which can be reproduced by a doublet triplet two-level model with an energy gap of 12 K.
These results suggest a non-magnetic $\Gamma_3$ doublet ground state of the Pr$^{3+}$ ion in the cubic crystalline electric field.
$C(T)$ exhibits a peak at $T_{\rm O}$ = 1.0 K and this peak remains robust against magnetic fields up to 5 T.
In powder neutron diffraction measurements, no magnetic reflection was observed at 0.32 K $<$ $T_{\rm O}$.
Two anomalies at $B$ = 2.1 and 5.3 T in magnetoresistance $\rho(B)$ at 0.05 K likely originate from switching in the order parameter.
These results suggest that the phase transition at $T_{\rm O}$ is ascribed to an antiferro-type order of the electric quadrupole or magnetic octupole of the $\Gamma_3$ doublet in the 4$f^2$ fcc lattice.
\end{abstract}

\maketitle

\section{Introduction}


Praseodymium-based cubic compounds with a 4$f^{2}$ configuration have attracted significant interest \cite{Morin90,Cox98,Sato09,Pr1-2-20}.
In these compounds, a strong interaction between their active multipolar degrees of freedom in the (quasi-) degenerated crystalline electric field (CEF) ground states and conduction electrons leads to various phenomena such as metal-insulator transition \cite{Sekine97,Iwasa05}, heavy-fermion superconductivity \cite{Bauer02,Kuwahara05,Yogi06}, and non-Fermi liquid (NFL) behavior due to the two-channel (quadrupole) Kondo effect \cite{Onimaru16,Yamane18,Tsuruta15}.

When the Pr$^{3+}$ ion site has a cubic point group, the CEF ground state can be a non-Kramers doublet. In this state, the magnetic dipole is quenched, and instead, the electric quadrupole and magnetic octupole become active. As a result, the quadrupole or octupole order may manifest itself \cite{Hattori14,Hattori16,Ishitobi21,Patri19}, while the two-channel Kondo effect is expected to occur if the quadrupoles are over-screened by equivalent conduction bands \cite{Affleck91,Jarrell96,Tsuruta99,Tsuruta15}.
An example is PrPb$_3$, which has a non-Kramers $\Gamma_{3}$ doublet ground state, showing an antiferroquadrupole (AFQ) order at $T_{\rm Q}$ $=$ 0.4 K \cite{Bucher72,Morin82,Tayama01}.
The order parameter is the $O_{2}^{0}$-type quadrupole \cite{Onimaru04}. 
Neutron diffraction measurements in a magnetic field along the [100] direction revealed incommensurate sinusoidal modulation of the quadrupoles \cite{Onimaru05}.
This modulated structure suggests not only long-range quadrupole interaction mediated by conduction electrons but also the quenching of the quadrupoles due to hybridization with the conduction electrons.
On the other hand, no long-range order of quadrupoles has been found in PrAg$_2$In and PrMg$_3$ with the $\Gamma_{3}$ doublet ground state. 
In these compounds, quadrupoles are prevented from ordering by either the quadrupole Kondo effect or atomic disorder inherent in their Heusler-type structures which lowers the site symmetry of Pr$^{3+}$~\cite{Yatskar96,Tanida06,Morie09}.

The discovery of the coexistence of the superconductivity and AFQ order in PrIr$_{2}$Zn$_{20}$ has drawn significant attention to  Pr 1-2-20 systems crystallizing in the cubic CeCr$_{2}$Al$_{20}$-type structure \cite{Nasch97}.
PrIr$_{2}$Zn$_{20}$ exhibits the AFQ order at $T_{\rm Q}$ $=$ 0.11 K below which the superconducting transition was observed at $T_{\rm c}$ $=$ 0.05 K \cite{Onimaru10,Onimaru11,Ishii11,Iwasa13,Iwasa17,Higemoto12,Woerl19,Kittaka24,Umeo20}.
The coexistence of the superconductivity and quadrupole order was also observed in isostructural PrRh$_{2}$Zn$_{20}$ \cite{Onimaru12,Yoshida17}, PrTi$_{2}$Al$_{20}$ \cite{Sakai11,Matsubayashi12,Sato12,Taniguchi16,Taniguchi20}, and PrV$_{2}$Al$_{20}$ \cite{Sakai11,Tsujimoto14}.
These findings imply that the superconductivity is related to the interaction between conduction electrons and the multipoles in the non-Kramers doublet \cite{Kubo20}.
Moreover, the NFL behaviors were observed not only in PrIr$_{2}$Zn$_{20}$ \cite{Onimaru16} and PrRh$_{2}$Zn$_{20}$ \cite{Yoshida17} but also in dilute Pr systems Y(Pr)Ir$_{2}$Zn$_{20}$ \cite{Yamane18,Yanagisawa19,Woerl22,Hibino23} and Y(Pr)Co$_{2}$Zn$_{20}$ \cite{Yamane20}.
These NFL behaviors probably result from the quadrupole Kondo effect to form
a composite electronic order involving the local quadrupole and the itinerant bands as the ground state \cite{Tsuruta15,Hoshino11,Hoshino13,Chandra13,Inui20}.

\begin{figure}[t]
\centering
\includegraphics[width=16pc]{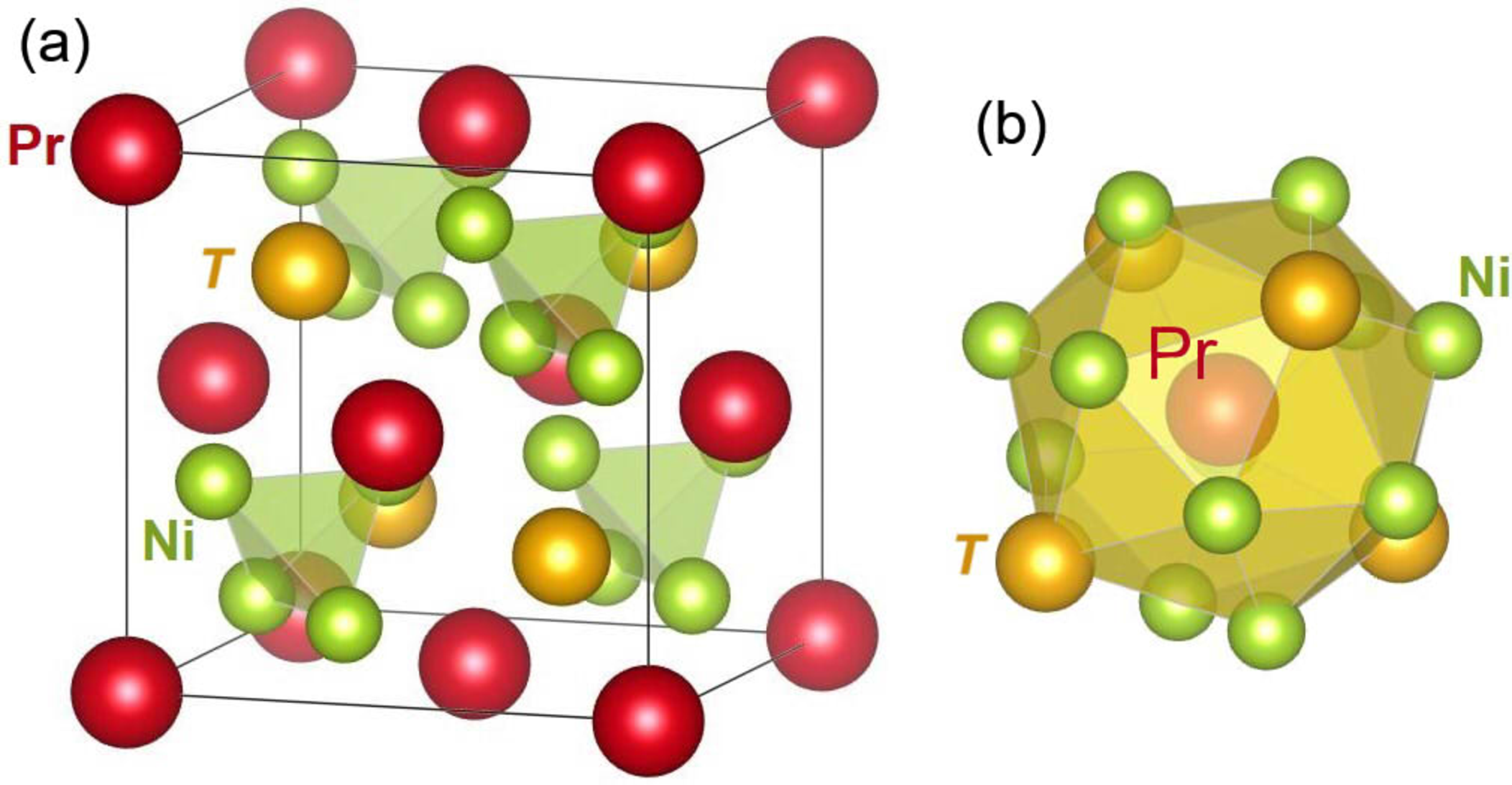}
\caption{(a) Cubic MgSnCu$_4$-type crystal structure of Pr$T$Ni$_4$  ($T$ = Mg, Cd, and In) with the space group of $F\bar{4}3m$~\cite{Kadir02}. 
In the structure, large (red) spheres represent Pr atoms, the medium (orange) spheres represent $T$ atoms, and the tetrahedra are formed by four Ni atoms shown with the small (green) spheres. 
The Pr sublattice forms an fcc lattice.
(b) A Pr-centered atomic cage made of four $T$ atoms at 4$c$ and 12 Ni atoms at 16$e$. 
The crystal structure images were drawn using VESTA \cite{Momma11}. 
}
\label{f1}
\end{figure}

Face-centered cubic (fcc) compounds Pr$T$Ni$_{4}$ ($T$ = Mg and In) are another family of Pr-based intermetallic compounds having the $\Gamma_{3}$ doublet ground state \cite{Kusanose19,Kusanose22,Tsujii02,Walker06}.
They crystallize in the cubic MgSnCu$_{4}$-type structure \cite{Kadir02}. As illustrated in Fig.~\ref{f1}(a),
since the Pr sublattice is equivalent to an fcc lattice, we anticipate an unusual ground state due to the geometrical frustration between active multipoles of the nearest neighbor Pr ions.
Additionally, anisotropic terms in the quadrupole interactions on the fcc lattice become effective alongside isotropic ones \cite{Tsunetsugu21, Hattori23}.
These potentials are quite distinct from the Pr-based compounds mentioned above in terms of the anisotropic and competitive multipolar interaction.
In fact, though PrInNi$_{4}$ exhibits ferromagnetic order at $T_{\rm C}$ = 0.75 K due to exchange interaction induced by coupling between the $\Gamma_{3}$ doublet and the first-excited triplet \cite{Tsujii02, Walker06}, PrMgNi$_4$ exhibits no phase transition down to 0.1 K \cite{Kusanose19}.
This hindered quadrupole order in PrMgNi$_4$ was attributed to a symmetry lowering caused by excess Mg atoms substituting for the Pr atoms or strong hybridizations between the 4$f^{2}$ electrons and conduction bands \cite{Kusanose19}. 
Otherwise, if the isotropic and anisotropic interactions in the fcc lattice are relatively weak compared to the energy splitting from the $\Gamma_{3}$ doubet to the excited $\Gamma_{1}$ singlet, the quadrupole order may be suppressed \cite{Tsunetsugu21, Hattori23}.

In our current study, we have focused on PrCdNi$_{4}$ crystallizing in the cubic MgSnCu$_{4}$-type structure \cite{Tappe12}.
As shown in Fig.~\ref{f1}(b), Pr is surrounded by Cd atoms at 4$c$ site and Ni atoms at 16$e$.
If the atomic disorder is reduced compared to the sister compound PrMgNi$_4$ \cite{Kusanose19}, the electric quadrupolar and/or magnetic octupolar order of the non-Kramers doublet may not be hindered.
With bearing this in mind, we conducted measurements of the transport and magnetic properties of PrCdNi$_{4}$ to determine the CEF ground state and understand the possible involvement of quadrupole and/or octupole in the formation of the ground state.
Powder neutron diffraction measurements were performed to judge whether the phase transition results from a magnetic or non-magnetic origin.
Recently, physical properties of the series of $R$CdNi$_4$ have been reported for $R$ = Ce, Nd, Sm, and Gd-Tm~\cite{Lee23} except for $R$ = Pr.

\section{Experimental Procedure}

\subsection{Preparation and characterization of samples}

We synthesized samples of PrCdNi$_4$ using the Cd self-flux method. 
First, we prepared the binary PrNi$_4$ alloy by arc melting.
Substantially, we sealed the PrNi$_4$ ingot and Cd shots in double quartz amples in an argon atmosphere.
The ampoules was then heated up to 1100$^{\circ}$C in an electric furnace and slowly cooled down.
At 500$^{\circ}$C, the ampoule was quickly removed from the furnace and centrifuged to remove the molten Cd flux.
The PrCdNi$_4$ samples were cubes of typically 3.0 mm$^3$ found to consist of grains smaller than 0.3 mm, as characterized by metallographic examination. 
The grain size was too small to select single crystalline samples for our transport and magnetic measurements.

The samples were analyzed using powder x-ray diffraction and electron-probe microanalysis (EPMA).
Backscattered electron images and x-ray diffraction patterns are shown in the Supplemental Material~\cite{supplement}.
The atomic compositions were determined by averaging over 10 different regions for each sample with a JEOL JXA-8200 analyzer.
Assuming that the sum of the compositions of the Pr and Cd atoms were 2, the atomic ratio of the sample batch for measurements was determined as Pr$_{1.00(1)}$Cd$_{1.00(1)}$Ni$_{3.89(4)}$, where the numbers in parentheses are the standard deviations.
It is noted that the composition obtained for Pr and Cd closely matches the stoichiometric ratio, which remains consistent across samples within the standard deviations.
Small amounts of impurity phases of PrNi$_2$Cd$_{20}$ and PrNi$_5$ were detected not only in the backscattered electron images but also in the powder x-ray diffraction patterns. 
The powder x-ray diffraction pattern confirms the cubic MgSnCu$_4$-type structure for the main phase.
The single-crystal x-ray structural analysis was performed at 293 K with a crystal smaller than 0.2 mm using the Mo $K{\alpha}$ radiation with the wavelength of $\lambda$ $=$ 0.71073 {\AA}, monochromated by a multilayered confocal mirror using a Rigaku XtaLAB Synergy-DW area-detector diffractometer.
Details of the measurement, data collection, and refinement are described in the Supplemental Material \cite{supplement}. 
Figure~\ref{f1}(a) shows the cubic MgSnCu$_4$-type structure of PrCdNi$_4$ \cite{Tappe12}.
In this structure, the Pr atoms occupy the fcc position of the unit cell, and the point group of the Pr site is the cubic $T_d$.
The structural refinement did not show any evidence for the site exchange between Pr and Cd sites.
The lattice constant was evaluated as $a$ $=$ 7.12932(8) {\AA} for PrCdNi$_{4}$. This value is 0.15\% larger than 7.11832(16) {\AA} for PrMgNi$_{4}$~\cite{Kusanose19}, determined by refining the powder x-ray diffraction pattern at room temperature.

\subsection{Physical property measurements}
The electrical resistance was measured using a standard four-probe AC method. 
The measurements were done with a Gifford-McMahon-type refrigerator for 3--300 K and with a commercial Cambridge Magnetic Refrigerator mFridge mF-ADR50 for 0.1--4 K.
The magnetic field dependences of $\rho(B)$ were measured up to 17.5 T at temperatures down to 0.05 K by the AC method using a $^3$He-$^4$He dilution refrigerator equipped with a 20 T magnet at NIMS.
Magnetization was measured from 1.8 to 300 K in magnetic fields for $B$ $\le$ 5 T using a commercial superconducting quantum interference device (SQUID) magnetometer (Quantum Design, MPMS). 
For 0.3 $<$ $T$ $<$ 4.2 K in $B \leq$ 8.5 T, a capacitive Faraday method was adopted.
Thereby, we used a high-resolution capacitive force-sensing device installed in a $^3$He single-shot refrigerator (Heliox, Oxford Instruments) \cite{Sakakibara94}.
The specific heat was measured using the thermal relaxation method in the temperature range of 0.4 $<$ $T$ $<$ 300 K for $B \leq$ 7 T with a Quantum Design physical property measurement system (PPMS).

In order to observe possible magnetic reflections in an ordered phase, powder neutron diffraction experiments were conducted using the ISSP triple-axis spectrometer GPTAS at JRR-3M in JAEA at Tokai, Japan~\cite{Nawa24}.
Neutron beams with a wavelength of $\lambda$ $=$ 2.4563 {\AA} ($\sim$13.7 meV) were obtained by the 002 reflection of a pyrolytic graphite (PG) monochromator. 
Horizontal collimation of 40$'$-Monochromator-40$'$-Sample-40$'$-Analyzer-80$'$ for a triple-axis mode was utilized.
A $^3$He refrigerator achieved the base temperature at 0.32 K. 
In order to prevent strong neutron absorption of Cd atoms, the powdered sample of 0.35 g was thinly spread (approximately 0.2 mm thick) on a single-crystalline silicon wafer~\cite{Ryan08, Tamura21}.
The scattering plane was tilted by 1 degree away from the [100] direction, which is perpendicular to the surface of the silicon wafer.

\section{Results and Discussion}

\begin{figure}
\centering
\includegraphics[width=20pc]{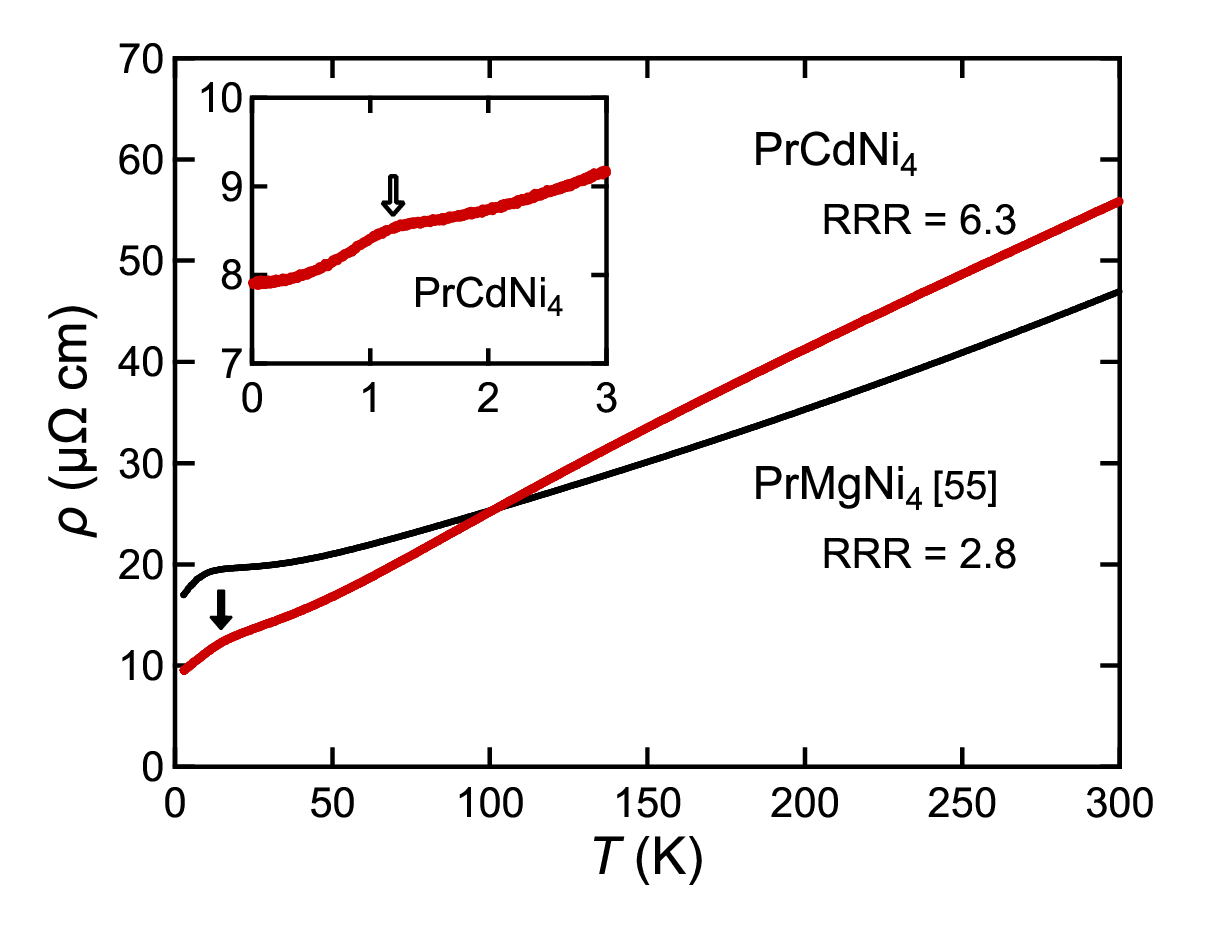}
\caption{Temperature dependence of the electrical resistivity $\rho(T)$ of PrCdNi$_4$ (red) and PrMgNi$_4$ (black)~\cite{Kusanose19}.
The inset displays the $\rho(T)$ data for $T$ $\le$ 3 K. The open arrow indicates a broad shoulder at 1.2 K.}
\label{fig02}
\end{figure}

\subsection{Electrical resistivity}

The temperature-dependent electrical resistivity $\rho(T)$ of PrCdNi$_4$ is compared with that of PrMgNi$_4$ \cite{Kusanose19} in Fig.~\ref{fig02}.
The residual resistivity ratio (RRR), evaluated as $\rho$(300~K)$/$$\rho$(0.1~K), is 6.3 for PrCdNi$_4$.
This RRR value is more than two times higher than 2.8 for a single crystalline PrMgNi$_4$ \cite{Kusanose19}.
The higher RRR value for PrCdNi$_4$ is consistent with the EPMA result.
The composition of Pr:Cd:Ni $=$ 1.00(1):1.00(1):3.89(4) is close to the stoichiometric ratio by assuming that the total composition of Pr and Cd is 2.
For the PrMgNi$_4$ sample, on the other hand, the composition was Pr:Mg:Ni $=$ 0.94(1):1.06(1):3.86(2), where excess Mg atoms substitute for the Pr atoms, leading to atomic disorder \cite{Kusanose19}.

As shown with the arrow in the main panel of Fig.~\ref{fig02}, $\rho(T)$ for PrCdNi$_4$ exhibits a shoulder at around 15 K, possibly due to the increase in scattering of conduction electrons by thermal excitations of the 4$f^2$ electrons between the CEF levels \cite{Abou75}.
This energy scale is consistent with the 12 K energy gap between the ground state $\Gamma_3$ doublet and the first excited triplet, as discussed later. 
Focusing on the lower temperature range in the inset of Fig.~\ref{fig02}, there is a broad shoulder centered at around 1.2 K.
This shoulder results from a phase transition observed in the specific heat measurements shown later.

\subsection{Magnetic susceptibility and isothermal magnetization}

\begin{figure}
\centering
\includegraphics[width=19pc]{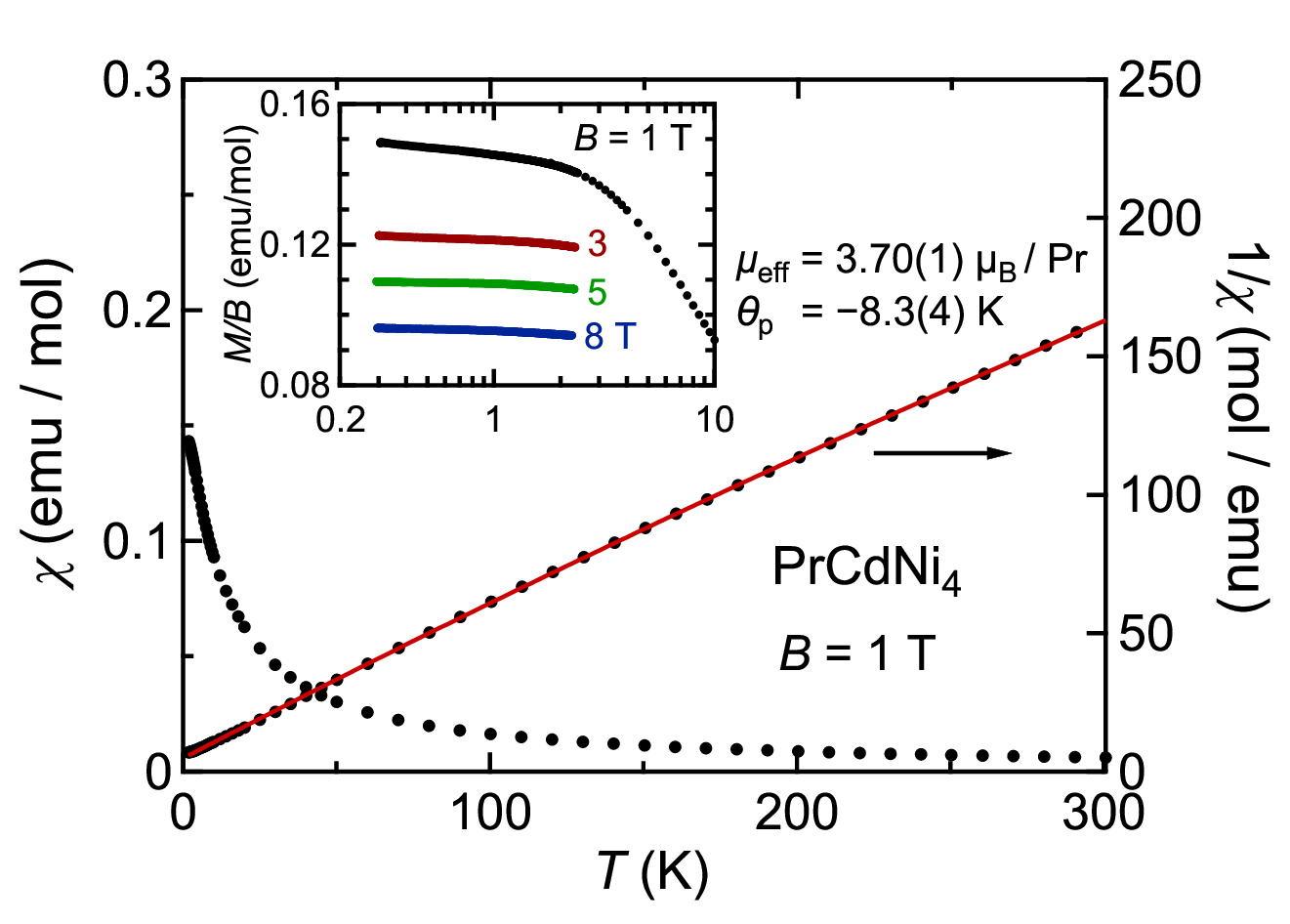}
\caption{Temperature dependence of the magnetic susceptibility, $\chi(T)$, and the inverse $\chi^{-1}(T)$ of PrCdNi$_4$ measured in a magnetic field of $B$ $=$ 1 T. 
The $\chi^{-1}$ data can be fitted with a modified Curie--Weiss equation. See text for details.
Inset shows the magnetization divided by the magnetic field, $M(T)/B$, at magnetic fields of $B$ $=$ 1, 3, 5, and 8 T, without any offset.
}
\label{fig03}
\end{figure}

Figure~\ref{fig03} shows the temperature dependence of the magnetic susceptibility $\chi(T)$ and the inverse $\chi^{-1}(T)$ of PrCdNi$_4$ measured in the magnetic field of $B$ $=$ 1 T. 
$\chi^{-1}(T)$ follows a modified Curie--Weiss equation: $\chi(T)$ $=$ $C$$/$($T$ $-$ $\theta_{\rm p}$) $+$ $\chi_{0}$, where $C$, $\theta_{\rm p}$, and $\chi_{0}$ represent the Curie constant, paramagnetic Curie temperature, and temperature independent susceptibility, respectively.
The (red) solid line represents the fit to the $\chi(T)$ data between 20 and 300 K using the above equation with $\theta_{\rm p}$ $=$ $-$8.3(4) K and $\chi_{0}$ $=$ 5.7(2) $\times$10$^{-4}$ emu/mol.
The negative value of $\theta_{\rm p}$ indicates antiferromagnetic intersite interaction between the Pr moments. 
The effective magnetic moment was evaluated to be $\mu_{\rm eff}$ $=$ 3.70(1) $\mu_{\rm B}$/f.u., which is moderately close to the value of 3.58 $\mu_{\rm B}$ for a free trivalent Pr ion. 
On cooling below 4 K, $\chi(T)$ does not diverge but approaches a constant, as shown in the inset.
This is a van-Vleck paramagnetic behavior of a nonmagnetic CEF ground state of the Pr ion.
As shown in the inset of Fig.~\ref{fig03}, the $M(T){/}B$ data at $B =$ 1, 3, 5, and 8 T do not exhibit a clear anomaly for $T$ $<$ 2 K.
%
\begin{figure}[b]
\centering
\includegraphics[width=20pc]{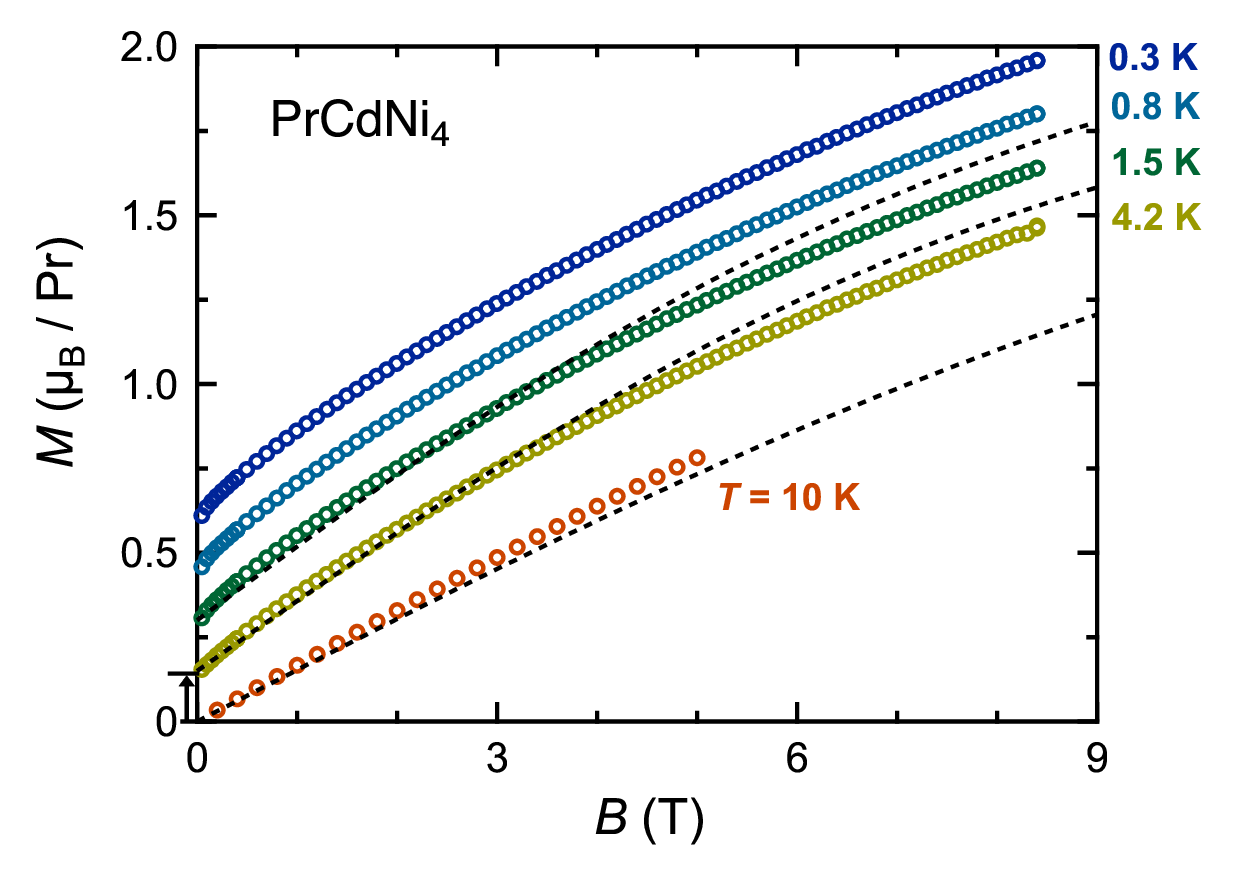}
\caption{Isothermal magnetization $M(B)$ of PrCdNi$_4$ at $T =$ 0.3, 0.8, 1.5, 4.2, and 10 K.
The data at $T \leq$ 4.2 K are vertically offset for clarity.
The dashed curves are calculations with the CEF parameters of $W$ $=$ $-$3.3 K and $x$ $=$ 0.8 and antiferromagnetic exchange interaction of $K_{1}$ $=$ $-$0.6 K between the Pr moments.
See text for details.
}
\label{fig04}
\end{figure}




The isothermal magnetization $M(B)$ data of PrCdNi$_4$ at temperatures of 0.3, 0.8, 1.5, 4.2, and 10 K are shown in Fig.~\ref{fig04}.
The data for temperatures below 4.2 K were obtained by the capacitive Faraday method, while the data at 10 K were measured by the SQUID magnetometer. 
The  $M(B)$ data for $T$ $\leq$ 4.2 K are vertically offset for clarity.
All $M(B)$ data show a monotonous increase with increasing magnetic field up to 8.5 T.
These magnetization curves are reproduced reasonably well using a CEF level scheme and antiferromagnetic exchange interaction, which will be discussed later.


\subsection{Specific heat and magnetic entropy}

\begin{figure}[b]
\centering
\includegraphics[width=21pc]{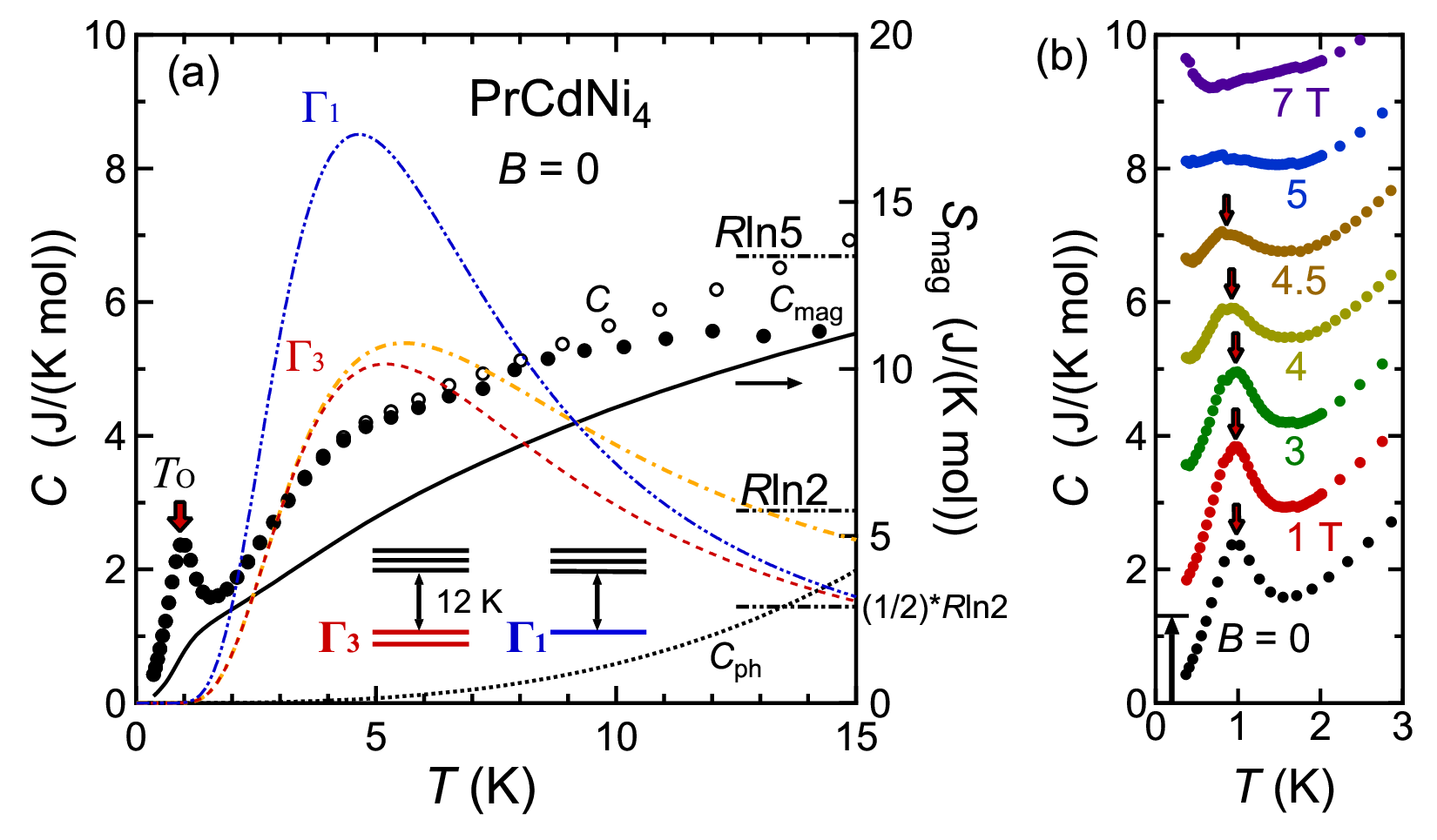}
\caption{(a) Temperature variations of the specific heat $C(T)$ (open circles) and the magnetic contribution $C_{\rm mag}$ (close circles), as well as the magnetic entropy $S_{\rm mag}$ (right-hand scale) of PrCdNi$_4$.
$C_{\rm ph}$ is the phonon contribution evaluated using a Debye model, as explained in the Supplemental Material~\cite{supplement}.
The (red) dashed and (blue) dotted curves represent the Schottky specific heat calculated by a two-level model of doublet triplet ($\Gamma_3$) and singlet triplet ($\Gamma_1$) with an energy gap of 12 K. 
The (orange) dot--dashed line represents the calculation with $W = -3.3$ K and $x$ $=$ 0.8 determined for PrMgNi$_{4}$ \cite{Kusanose22}.
(b) The specific heat $C(T)$ at magnetic fields of $B \leq $ 7 T. 
The data for $B$ $\ge$ 1 T are offset for clarity.
The arrows indicate the peak temperature, which remains robust against magnetic fields up to 4.5 T.
}
\label{fig05}
\end{figure}

The specific heat $C(T)$ data of PrCdNi$_4$ for $T$ $\le$ 15 K are shown in Fig.~\ref{fig05}(a). 
Since a nonmagnetic counterpart LaCdNi$_4$ could not be synthesized, the phonon contribution $C_{\rm ph}(T)$ was evaluated using the Debye model with $\theta_{\rm D}$ $=$ 270.7(6) K as described in the Supplemental Material \cite{supplement,Kittel05}.
By subtracting $C_{\rm ph}(T)$ from the total $C(T)$ data, we estimated the magnetic contribution $C_{\rm mag}(T)$.
$C_{\rm mag}$ exhibits a shoulder at around 4 K, which is attributable to the Schottky specific heat due to thermal excitations from the CEF ground state to excited levels.

A two-level Schottky model provides the expression of $C(T)$ with the equation as
\begin{equation}
	C = \frac{n m \Delta^2}{k_{\rm B} T^2}  \frac{\mathrm{e}^{- {\Delta} / {k_{\rm B} T}}}{ \left( n + m \mathrm{e}^{- {\Delta} / {k_{\rm B} T}} \right) ^2} \label{C_Sch},
\end{equation}
where $n$ and $m$ represent the degeneracy of the ground and excited multiplets, respectively, and $\Delta$ is the energy gap between the two levels.
Considering the van Vleck paramagnetic behavior shown in the inset of Fig.~\ref{fig03}, the CEF ground state of the Pr ion for the cubic $T_d$ point group can be either the nonmagnetic $\Gamma_{1}$ singlet or $\Gamma_{3}$ doublet.
Calculations with a doublet triplet ($\Gamma_{3}$) two-level model and a singlet triplet ($\Gamma_{1}$) one, with an energy split of 12 K, are shown with the (red) dashed and (blue) dotted curves, respectively.
It is assumed that the first-excited state is a triplet, i.e., either the $\Gamma_{4}$ or $\Gamma_{5}$ triplet \cite{Lea62}.
The respective two-level schemes are depicted in the lower part of Fig.~\ref{fig05}(a).
The $\Gamma_{3}$ model better reproduces the shoulder of $C_{\rm mag}(T)$ at around 4 K than the $\Gamma_{1}$ model. 
Therefore, the CEF scheme likely consists of the $\Gamma_{3}$ doublet ground state and the first-excited triplet.
Note that the absolute value of $C_{\rm mag}$ at around 4 K is smaller than that expected by the $\Gamma_{3}$ model.
This discrepancy may be ascribed to 
a relatively small fraction of the pristine PrCdNi$_4$ in the measured sample, which could contain some impurities of PrNi$_5$ and PrNi$_2$Cd$_{20}$.
On the other hand, the larger value of $C_{\rm mag}$ observed above 7 K may result from the contribution of the CEF levels at higher energy. 
Otherwise, it is attributable to an underestimation of $C_{\rm ph}$.

In order to verify this CEF scheme, we calculated the isothermal magnetization.
The cubic CEF Hamiltonian is described by the equation as
\begin{equation*}
\mathcal{H}_{\rm CEF}=W \left[ \frac{x}{60} \left( O_{4}^{0}+5O_{4}^{4} \right) +\frac{1-|x|}{1260} \left( O_{6}^{0}-21O_{6}^{4} \right) \right],
\end{equation*}
where $W$ and $x$ represent the CEF parameters, and $O_{n}^{m}$'s stand for the Stevens operators \cite{Lea62}. 
The energy gap of 12 K between the $\Gamma_{3}$ doublet and the first excited triplet is nearly the same as that observed in PrMgNi$_{4}$ \cite{Kusanose19}, although it is anticipated that the energy splitting would be smaller in PrCdNi$_{4}$ due to its larger lattice parameter.
Therefore, we use the parameters of $W$ $=$ $-$3.3 K and $x$ $=$ 0.8 for PrMgNi$_{4}$ as determined by the inelastic neutron scattering experiments; $\Gamma_{3}$(0)--$\Gamma_{4}$(13.2 K)--$\Gamma_{1}$(31.7 K)--$\Gamma_{5}$(134.6 K) \cite{Kusanose22}.
The isothermal magnetization is calculated with the above CEF parameters and an intersite magnetic interaction using the following Hamiltonian.
\begin{equation*}
\mathcal{H} = \mathcal{H}_{\rm CEF} + g_{J}\mu_{\rm B}\textbf{\textit J}\textbf{\textit B} - {K_{1}} \langle\textbf{\textit J}\rangle\textbf{\textit J},
\end{equation*}
where $g_{J}=$  4$/$5 is the Land$\acute{\rm e}$ $g$ factor for a Pr$^{3+}$ ion, \textit{\textbf J} is a total angular momentum, and $K_{1}$ is a coefficient of the magnetic inter site interaction between the Pr ions.
The isothermal magnetization calculated  for \textit{\textbf B} $||$ [111] with antiferromagnetic interaction of $K_{1}$ $=$ $-$0.6 K is shown with the dashed lines in Fig.~\ref{fig04}.
The $M(B)$ calculations at $T$ $=$ 1.5, 4.2, and 10 K match the data of the polycrystalline sample.
It is noted that the specific heat calculated using the above CEF parameters moderately reproduces the shoulder of $C_{\rm mag}$ around 4 K, as depicted by the (orange) dot-dashed line in Fig.~\ref{fig05}. These results give further support that the CEF ground state is the $\Gamma_{3}$ doublet carrying the electric quadrupole and magnetic octupole.

With decreasing temperature below 2 K, $C_{\rm mag}(T)$ exhibits a peak at $T_{\rm O}$ $=$ 1.0 K. 
This peak is the manifestation of a phase transition of the $\Gamma_{3}$ doublet.
It is crucial to exclude the contribution of the impurity phases; neither PrNi$_5$ nor PrNi$_2$Cd$_{20}$ exhibit any phase transition near 1 K~\cite{Ott76, Sebek00, Burnett14, Yanagisawa15}.
The magnetic entropy $S_{\rm mag}$, estimated by integrating the $C_{\rm mag}{/}T$ data with respect to temperature, is plotted on the right-hand axis of Fig.~\ref{fig05}(a).
Here, we assumed that the $C_{\rm mag}/{T}$ would approach zero linearly as $T$ decreases from 0.4 to 0 K. 
Then, we estimated the magnetic entropy $S_{\rm mag}$ at 0.4 K to be 0.21 J/K, which was added to $S_{\rm mag}$($T$) for $T >$ 0.4 K.
At $T_{\rm O}$, $S_{\rm mag}$ is only 40\% of $R$ln2, and at around 6 K it reaches $R$ln2 expected for the full order of the doublet.
This suggests that the phase transition is due to the multipolar degrees of freedom in the $\Gamma_{3}$ doublet, such as $\Gamma_{3}$-type electric quadrupoles or $\Gamma_{2}$-type magnetic octupole.
$S_{\rm mag}$ reaches $R$ln5 at 18 K, which is consistent with the aforementioned CEF level scheme with the triplet state separated by 12 K from the $\Gamma_3$ doublet ground state.

Figure~\ref{fig05}(b) shows the $C(T)$ data in magnetic fields of $B \leq$ 7 T.
The peak temperature $T_{\rm O}$ remains unchanged until the magnetic field is increased up to 4.5 T.
The robustness of $T_{\rm O}$ to the magnetic field is reasonable when the phase transition does not arise from the magnetic dipole but from the quadrupole or the octupole in the $\Gamma_{3}$ doublet ground state.
With further increase in the magnetic field above 5 T, the peak is suppressed and becomes vague.

\begin{figure}
\centering
\includegraphics[width=18pc]{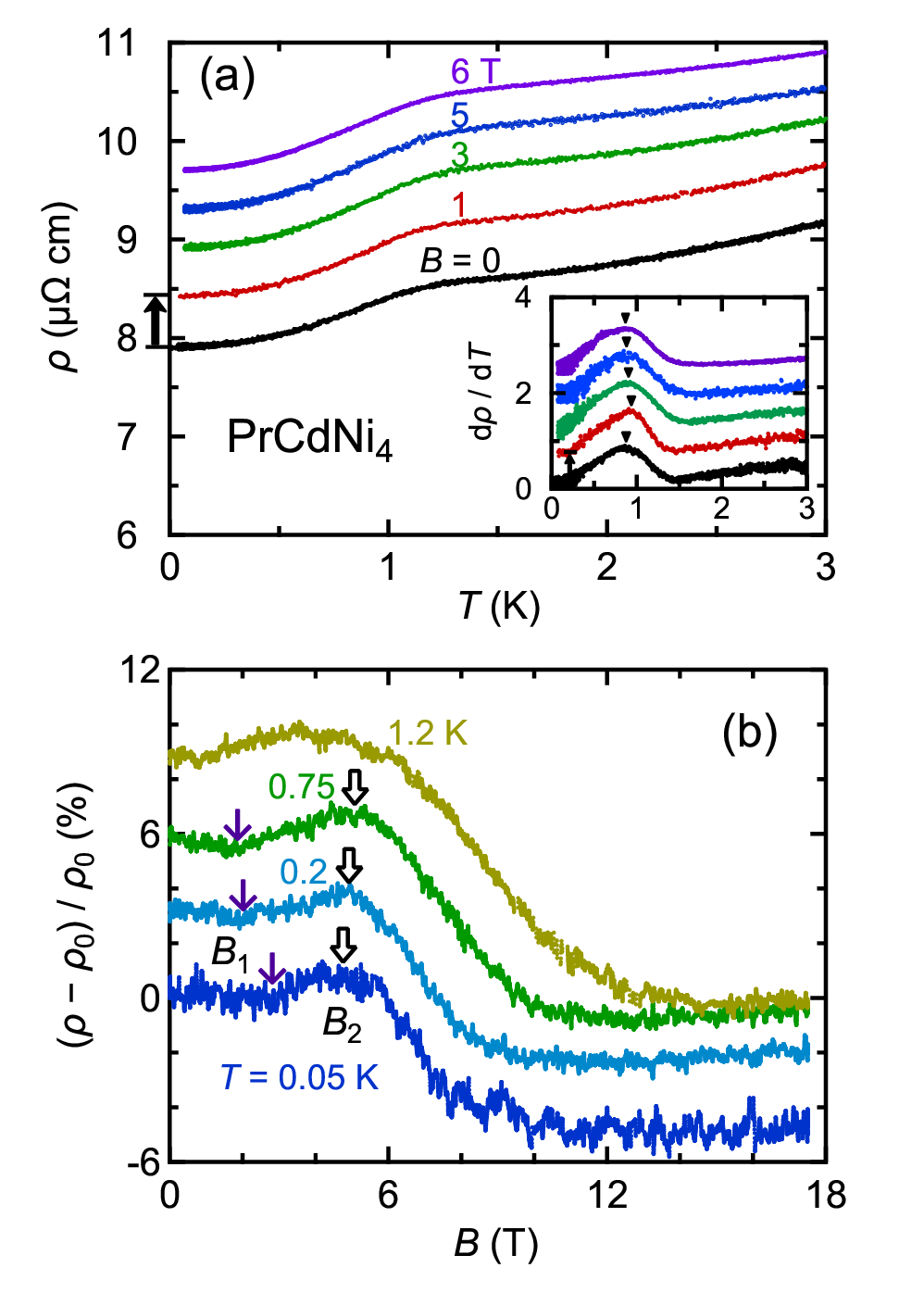}
\caption{(a) Temperature dependence of the electrical resistivity $\rho(T)$ of PrCdNi$_{4}$ for $T$ $<$ 3 K in various constant magnetic fields $B$ up to 6 T.
Inset shows the derivative $d\rho/dT$, where the arrows indicate the peak temperatures. 
The data in $B$ are offset for clarity; for example, the bold arrow outside the vertical left axis indicates the offset for the data at $B$ $=$ 1 T.
(b) Normalized magnetoresistance at various constant temperatures 0.05, 0.2, 0.75, and 1.2 K in $B$ up to 17.5 T.
The arrows highlight a shallow minimum at 2.8 T and a hump at 4.7 T.
}
\label{fig06}
\end{figure}

\subsection{Electrical resistivity in magnetic fields}

Figure~\ref{fig06}(a) shows the electrical resistivity $\rho(T)$ in magnetic fields of $B$ $=$ 0, 1, 3, 5, and 6 T.
Here, the direction of the magnetic field is parallel to that of the electrical current.
In the inset of Fig. \ref{fig06}(a), the derivative of the electrical resistivity, $d{\rho}(T)/dT$, is shown.
At $B$ $=$ 0, a broad maximum is observed at 0.88 K, which is comparable to $T_{\rm O}$, the peak of $C_{\rm mag}(T)$.
This maximum slightly shifts to lower temperatures with increasing magnetic fields, reaching 0.86 K at $B$ $=$ 6 T.
The magnetic-field variation of the maximum is consistent with that of $T_{\rm O}$ in $C(T)$ shown in Fig. \ref{fig05}(b).

Figure~\ref{fig06}(b) shows the magnetic-field dependence of the normalized magnetoresistance $(\rho - \rho_{\rm 0}) /\rho_{\rm 0}$ at various constant temperatures.
The data were obtained in magnetic fields up to 17.5 T at temperatures of 0.05, 0.2, 0.75, and 1.2 K.
At 0.05 K, there is a shallow minimum at $B_{\rm 1}$ $=$ 2.8 T and a hump at $B_{\rm 2}$ $=$ 4.7 T.
At $B >$ 5 T, the magnetoresistance decreases and then remains constant for $B$ $\ge$ 10 T.
The anomalies at $B_{\rm 1}$ and $B_{\rm 2}$ exist at 0.2 and 0.75 K.
However, at 1.2 K above $T_{\rm O}$, the minimum at $B_{\rm 1}$ disappears, and the hump changes to a broad maximum.
These anomalies are probably attributed to the switching of the order parameter in the ordered state, as discussed later.

\begin{figure}
\centering
\includegraphics[width=21pc]{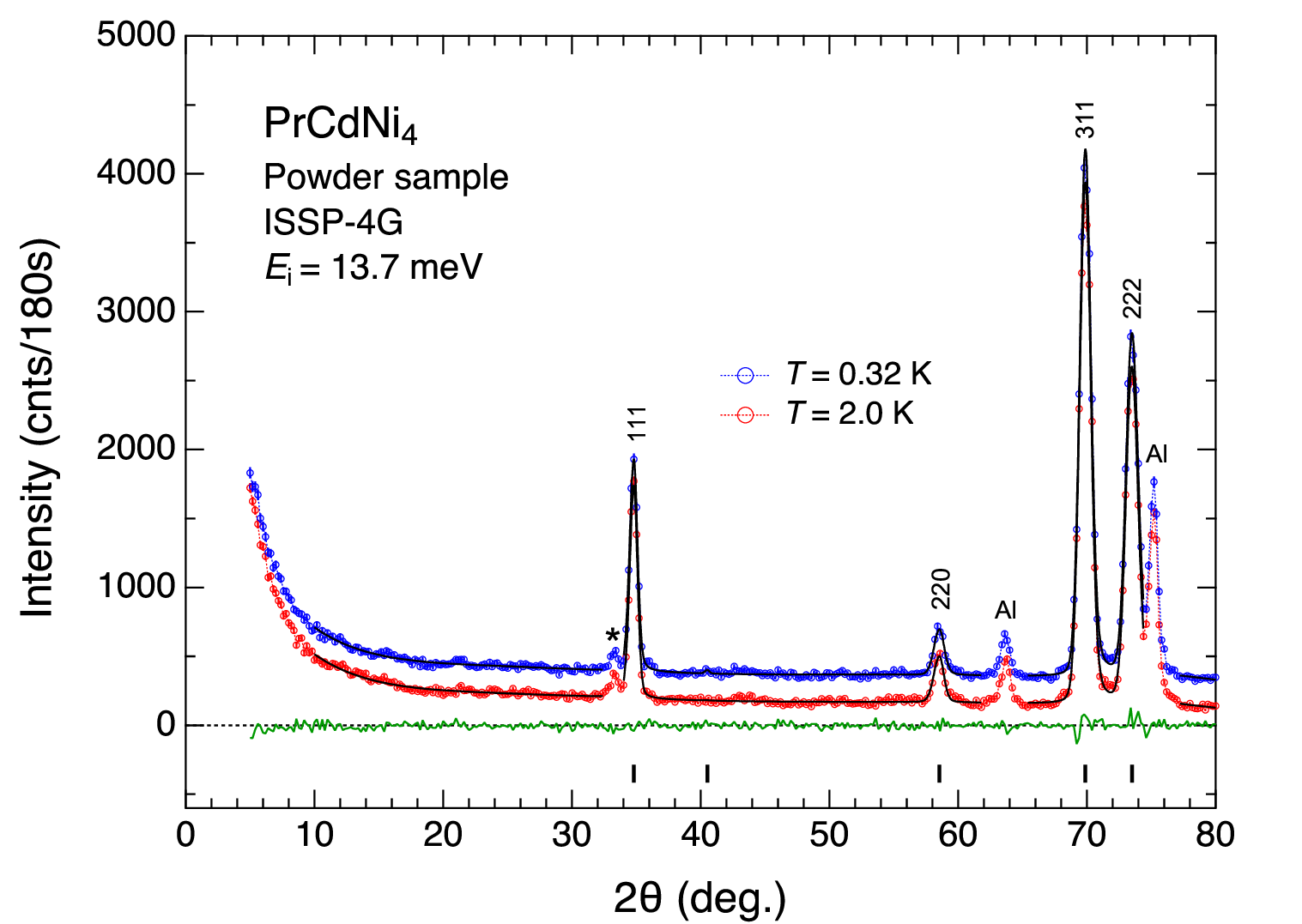}
\caption{Powder neutron diffraction patterns of PrCdNi$_4$ at $T$ $=$ 2.0 K (red) and 0.32 K (blue), measured above and below $T_{\rm O}$ $=$ 1.0 K, respectively.
The (green) line shows the difference between the data at 2.0 and 0.32 K.
The black lines represent the fits with the Rietveld analysis.
The bars indicate the scattering angles for the nuclear Bragg peaks of PrCdNi$_4$
A peak indicated by a star signifies an impurity phase of PrNi$_{5}$.
}
\label{fig07}
\end{figure}

\subsection{Powder neutron diffraction}

Neutron diffraction patterns of the powdered sample of PrCdNi$_4$ at $T$ $=$ 2.0 K (red) $> T_{\rm O}$ and 0.32 K (blue) $< T_{\rm O}$ are represented in Fig.~\ref{fig07}.
Most of the observed peaks are identified as nuclear peaks of PrCdNi$_4$ and aluminum from the sample container.
A peak at 2$\theta$ $=$ 33.2$^{\circ}$, indicated by a star, is ascribed to a small amount of impurity PrNi$_{5}$.
In the differential pattern (green), no peak appears.
Therefore, it is concluded that the phase transition at $T_{\rm O}$ does not result from a magnetic origin.
This finding is consistent with the manifestation of the multipole order in the $\Gamma_3$ doublet ground state, as will be discussed with the $B$--$T$ phase diagram.

The Rietveld profile fitting was conducted to determine the structural parameters of PrCdNi$_4$.
The fits are shown with the (black) solid lines in Fig. \ref{fig07}.
The lattice parameter of $a$ $=$ 7.1112(3) {\AA} at 0.32 K was estimated by RIETAN-FP~\cite{RIETAN}.
Details are provided in the Supplemental Materials \cite{supplement}.


\subsection{Magnetic field vs temperature phase diagram}

\begin{figure}
\centering
\includegraphics[width=16pc]{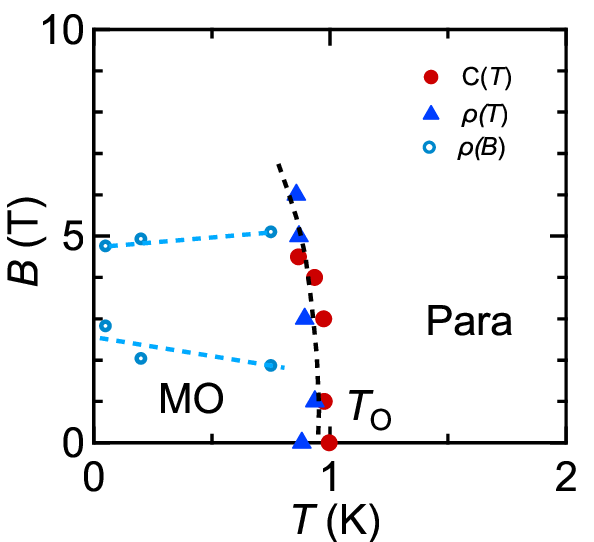}
\caption{Magnetic field ($B$) vs temperature ($T$) phase diagram of PrCdNi$_4$ determined by the measurements of the specific heat $C(T)$, electrical resistivity $\rho(T)$, and magnetoresistance $\rho(B)$. 
``MO" represents a multipole order, $T_{\rm o}$ is the ordering temperature, and ``Para" indicates the state where long-range multipole order is absent.
}
\label{fig08}
\end{figure}

Figure \ref{fig08} shows the magnetic-field vs temperature ($B$--$T$) phase diagram of PrCdNi$_4$.
The phase transition temperature $T_{\rm O}$ is determined from the $C(T)$ and $\rho(T)$ data.
$T_{\rm O}$ remains almost unchanged up to $B$ = 6.5 T.
The robustness of $T_{\rm O}$ to the magnetic field is a characteristic of the multipole order, such as the electric quadrupole or magnetic octupole order in the $\Gamma_3$ doublet ground state.
This behavior is consistent with 
the absence of magnetic reflection at 0.32 K.
Moreover, the anomalies found in the $\rho(B)$ data are plotted in the $B$--$T$ phase diagram.
Two horizontal lines at 2 and 5 T look like phase boundaries within the ordered phase.
Since $M(B)$ shows no anomaly at 0.3 K, as shown in Fig. 4, these boundaries must relate to the multipolar order parameter rather than the realignment of magnetic dipoles in a conventional antiferromagnetic order. 
Thereby, these boundaries are likely attributed to the switching of the order parameters as observed in the AFQ order in the Pr-based compounds with the $\Gamma_{3}$ doublet ground state \cite{Onimaru05,Onimaru11,Kittaka24}.
When ferroquadrupole (FQ) order arises from isotropic quadrupole interactions, its uniform order parameter remains unchanged in a magnetic field. 
Therefore, these results suggest that an antiferro-type multipole order manifests itself below $T_{\rm O}$. 

Let us discuss why the multipolar transition manifests itself in PrCdNi$_{4}$ even though the long-range order of the $\Gamma_{3}$ doublet is hindered in the isostructural PrMgNi$_{4}$ \cite{Kusanose19}. 
One possible reason is that the composition of Pr$_{1.00(1)}$Cd$_{1.00(1)}$Ni$_{3.89(4)}$ is close to the stoichiometric ratio, while excess Mg atoms substitute for the Pr atoms in the sample of Pr$_{0.94(1)}$Mg$_{1.06(1)}$Ni$_{3.86(2)}$ \cite{Kusanose19}.
Without the atomic disorder in PrCdNi$_{4}$, the non-Kramers $\Gamma_{3}$ doublet's degeneracy could be conserved to give rise to the phase transition due to intersite exchange interactions. 
It has been noted that the AFQ order easily collapses due to a small amount of atomic exchange \cite{Kawae01,Matsumoto15}. This may also apply to the absence of quadrupole order in PrMgNi$_{4}$, which further supports the AFQ order in PrCdNi$_{4}$.

On the other hand,  the competitive isotropic $J$ and anisotropic $K$ nearest-neighbor interactions in the fcc lattice may affect quadrupolar order, combined with the CEF effect, as theoretically investigated using a four-site mean-field approximation \cite{Tsunetsugu21}.
According to this calculation,
the order parameter is strongly dependent on the values of $J$ and $K$ scaled by the excitation energy from the $\Gamma_{3}$ doublet to the excited $\Gamma_{1}$ singlet, denoted as $E_{1}$.
Regarding the finite temperature properties of the quadrupole orders, there appear not only the FQ phase, as well as $O_2^0$- and $O_2^2$-type AFQ phases, but also triple-$q$ order with partially ordered sites. 
Moreover, the transition temperature is significantly suppressed over a wide range of small values of $J$ and $K$ \cite{Tsunetsugu21}.
In PrCdNi$_{4}$, $E_{1}$ is estimated as 31.7 K for $W = -$3.3 K and $x =$ 0.8, leading to $T_{\rm O}$$/$$E_{1}$ $=$ 0.03.
This value is roughly comparable to that obtained where the absolute values of $K$/$E_{1}$ and $J$/$E_{1}$ are less than 0.01 in the $J$-$K$ phase diagram.
Furthermore, since the phase transition is of second order and no successive transition was observed, the ground state is likely characterized by the $O_2^2$-type AFQ order.


To reveal the impact of the competitive anisotropic and isotropic interactions between the multipoles inherent in the fcc lattice, it is essential to observe the anisotropic dependence of physical properties with respect to the crystal axes and external field axes using single-crystalline samples.

\section{Summary}

We have conducted various measurements on a cubic compound PrCdNi$_4$, including magnetic susceptibility $\chi$, isothermal magnetization $M$, specific heat $C$, electrical resistivity $\rho$, magnetoresistance, and powder neutron diffraction.
$\chi(T)$ approaches a constant value on cooling below 10 K. 
This van-Vleck paramagnetic behavior is consistent with the expected non-magnetic CEF ground state for Pr$^{3+}$.
The magnetic specific heat $C_{\rm mag}$ exhibits a shoulder at around 4 K, ascribed to CEF excitations from the $\Gamma_3$ doublet ground state to an excited triplet.
Furthermore, $C(T)$ shows a peak  at $T_{\rm O}$ $=$ 1.0 K, which temperature does not significantly change in magnetic fields up to 4.5 T.
This peak is likely attributed to multipolar degrees of freedom in the $\Gamma_3$ doublet.
No magnetic reflections were observed below $T_{\rm O}$ in the powder neutron diffraction pattern.
It is noted that $S_{\rm mag}$ reaches only 40\% of $R$ln2 at $T_{\rm O}$. 
This reduction in the entropy is probably attributed to competitive anisotropic multipole interactions inherent in the 4$f^2$ fcc lattice~\cite{Tsunetsugu21}.
Moreover, the magnetoresistance $\rho(B)$ at $T$ $=$ 0.05 K ($<$ $T_{\rm O}$) showed
two anomalies at $B$ $=$ 2.8 and 4.7 T which are attributable to switching in the order parameter.
If this is the case, the phase transition at $T_{\rm O}$ results from an antiferro-type multipole order in the $\Gamma_3$ doublet on the $4f^{2}$ fcc lattice.
To further understand the order parameter, it is necessary to measure the anisotropy of the magnetic and transport properties using single-crystalline samples of PrCdNi$_4$.
In addition, neutron diffraction and nuclear magnetic/quadrupole resonance measurements should be conducted in different orientations of magnetic fields.
Resonant x-ray diffraction and ultrasonic measurements are promising methods for directly detecting quadrupoles and/or octupoles in zero magnetic field.
\\

\section*{Acknowledgments}

The authors would like to thank Y. Yamane, H. Kusunose, T. Ishitobi, K. Hattori, M. Nohara, T. Matsumura, H. Funashima, and H. Harima for helpful discussion. 
The authors also would like to thank R. Yamamoto and S. Mizutani for their measurements of the low-temperature magnetoresistance at NIMS.
The authors thank Y. Shibata for the electron-probe microanalysis carried out at N-BARD, Hiroshima University. 
The measurements of the magnetization with MPMS and the capacitance Faraday method with the $^3$He Heliox refrigerator, the specific heat with the PPMS and the Cambridge Magnetic Refrigerator mFridge mF-ADR50 were performed at N-BARD, Hiroshima University. 
The single-crystal x-ray structural analysis was performed using a Rigaku XtaLAB Synergy-DW area-detector diffractometer at N-BARD, Hiroshima University.
MANA is supported by World Premier International Research Center Initiative (WPI), MEXT, Japan.
This work was carried out by the JRR-3 general user program managed by the Institute for Solid State Physics, the University of Tokyo (Proposal No.~22514). 

This work was financially supported by MEXT/JSPS KAKENHI Grants No. JP26707017, No. JP15H05886 (J-Physics), No. JP18H01182, No. JP21J14456, No. JP22H00101, No. JP22K03529, No. JP23H04870, No. JP23KK0051 and No. JP24K00574, Japan. Additional support was provided by JST FOREST JPMJFR2233.

\end{document}


\title{Supplemental Materials for  \\
`{Multipolar Phase Transition in the 4$f^2$ fcc Lattice Compound PrCdNi$_{4}$}'
}

\author{Yuka Kusanose}
	\email{kusanose.yuka.p7@f.mail.nagoya-u.ac.jp}
	\affiliation{
	Department of Quantum Matter, Graduate School of Advanced Science and Engineering, Hiroshima University, Higashi-Hiroshima 739-8530, Japan}
	\affiliation{
	Department of Applied Physics, Nagoya University, Nagoya 464-8603, Japan}
\author{Yasuyuki Shimura}
	\affiliation{
	Department of Quantum Matter, 
	Graduate School of Advanced Science and Engineering, Hiroshima University, Higashi-Hiroshima 739-8530, Japan}
\author{Kazunori Umeo}
	\affiliation{Natural Science Center for Basic Research and Development (N-BARD),
	Hiroshima University, Higashi-Hiroshima 739-8526, Japan}
\author{Naomi Kawata}
	\affiliation{Natural Science Center for Basic Research and Development (N-BARD),
	Hiroshima University, Higashi-Hiroshima 739-8526, Japan}
\author{Toshiro Takabatake}
	\affiliation{
	Department of Quantum Matter, 
	Graduate School of Advanced Science and Engineering, Hiroshima University, Higashi-Hiroshima 739-8530, Japan}
\author{Taichi Terashima}
	\affiliation{
	Research Center for Materials Nanoarchitectonics (MANA), National Institute for Materials Science (NIMS), Tsukuba 305-0003, Japan}
\author{Naoki Kikugawa}
	\affiliation{
	Center for Basic Research on Materials (CBRM), National Institute for Materials Science (NIMS), Tsukuba 305-0003, Japan}
\author{Takako Konoike}
	\affiliation{
	Research Center for Materials Nanoarchitectonics (MANA), National Institute for Materials Science (NIMS), Tsukuba 305-0003, Japan}
\author{Yuya Hattori}
	\affiliation{
	Center for Basic Research on Materials (CBRM), National Institute for Materials Science (NIMS), Tsukuba 305-0003, Japan}
\author{Kazuhiro Nawa}
	\affiliation{
	Institute of Multidisciplinary Research for Advanced Materials, Tohoku University, Sendai 980-8577, Japan}
\author{Hung-Cheng Wu}
	\altaffiliation[Present address: ]{National Sun Yat-sen University}
	\affiliation{
	Institute of Multidisciplinary Research for Advanced Materials, Tohoku University, Sendai 980-8577, Japan}
\author{Taku J. Sato}
	\affiliation{
	Institute of Multidisciplinary Research for Advanced Materials, Tohoku University, Sendai 980-8577, Japan}
\author{Takahiro Onimaru}
	\affiliation{
	Department of Quantum Matter, 
	Graduate School of Advanced Science and Engineering, Hiroshima University, Higashi-Hiroshima 739-8530, Japan}

\date{\today}

\maketitle

\section{Sample characterization}

Backscattered electron images and powder x-ray diffraction patterns for samples A and B of polycrystalline PrCdNi$_{4}$ are shown in Figs. S1 and S2, respectively.
It is noted that a polycrystalline sample from batch A was selected because it contains multiple phases. 
This sample includes not only PrCdNi$_{4}$ but also impurity phases such as PrNi$_{2}$Cd$_{20}$ and PrNi$_{5}$, which is consistent with the emergence of additional peaks measured by the powder x-ray diffraction shown in Fig. S2. 
The atomice composition of the PrCdNi$_{4}$ phase was determined by electron-probe microanalysis (EPMA) to be Pr$_{0.99(1)}$Cd$_{1.01(1)}$Ni$_{3.87(3)}$. 
On the other hand, another sample from batch B is nearly single-phase, but it contains small amounts of PrNi$_{5}$ and Cd impurities.
As shown in Fig. S2, there are some peaks from PrNi$_{5}$ in the powder x-ray diffraction pattern, although the peaks from PrNi$_{2}$Cd$_{20}$ are less pronounced. 
The composition for the PrCdNi$_{4}$ phase of sample B was determined to be Pr$_{0.97(1)}$Cd$_{1.03(1)}$Ni$_{3.84(6)}$ by EPMA. 
The compositions of samples A and B are similar to Pr$_{1.00(1)}$Cd$_{1.00(1)}$Ni$_{3.89(4)}$ of the sample batch for measurements, and they are close to the stoichiometric ratio compared with Pr$_{0.94(1)}$Mg$_{1.06(1)}$Ni$_{3.86(2)}$ in Ref. \cite{Kusanose19}, which shows no phase transition.

\begin{figure}[H]
\centering
\includegraphics[width=30pc]{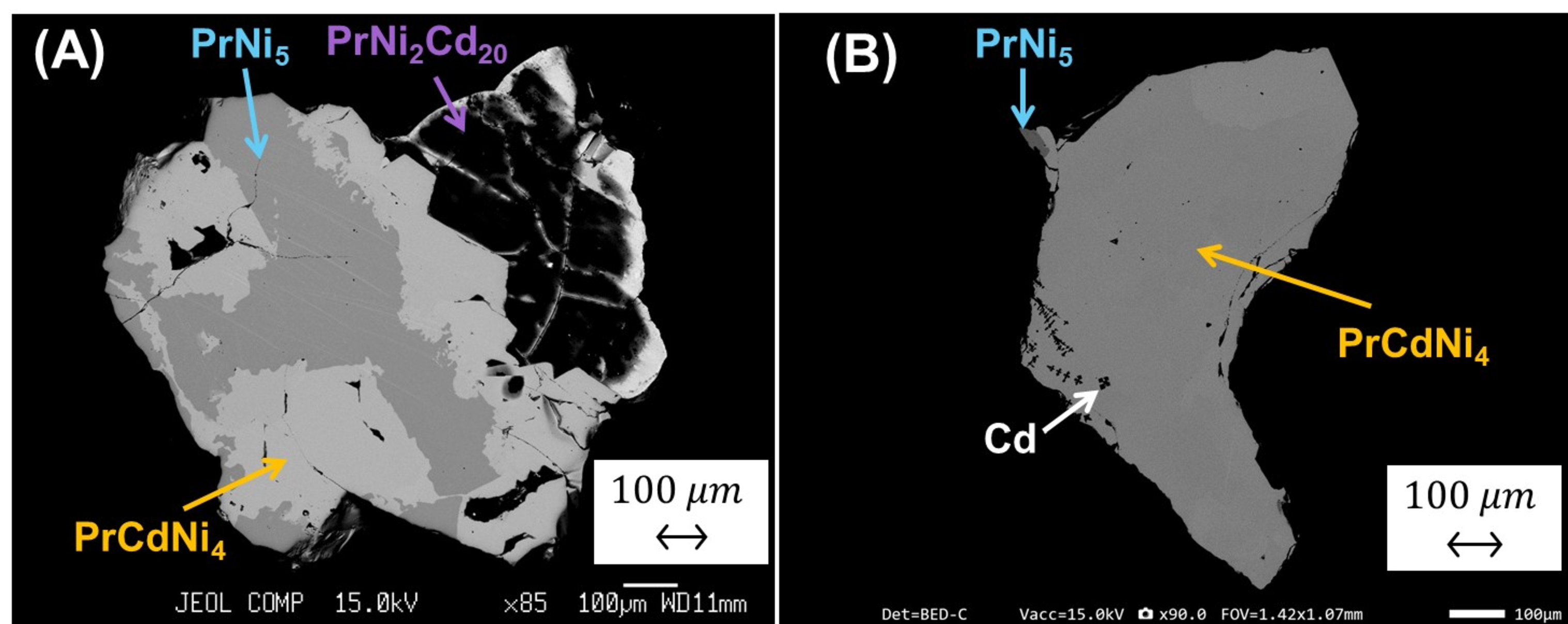}
\caption*{FIG. S1. Backscattered electron images of the samples A and B of PrCdNi$_4$.
Sample A contains PrCdNi$_{4}$ along with impurity phases, such as PrNi$_{2}$Cd$_{20}$ and PrNi$_{5}$, whereas sample B is nearly a single phase.
}
\label{FigS01}
\end{figure}
\begin{figure}[H]
\centering
\includegraphics[width=30pc]{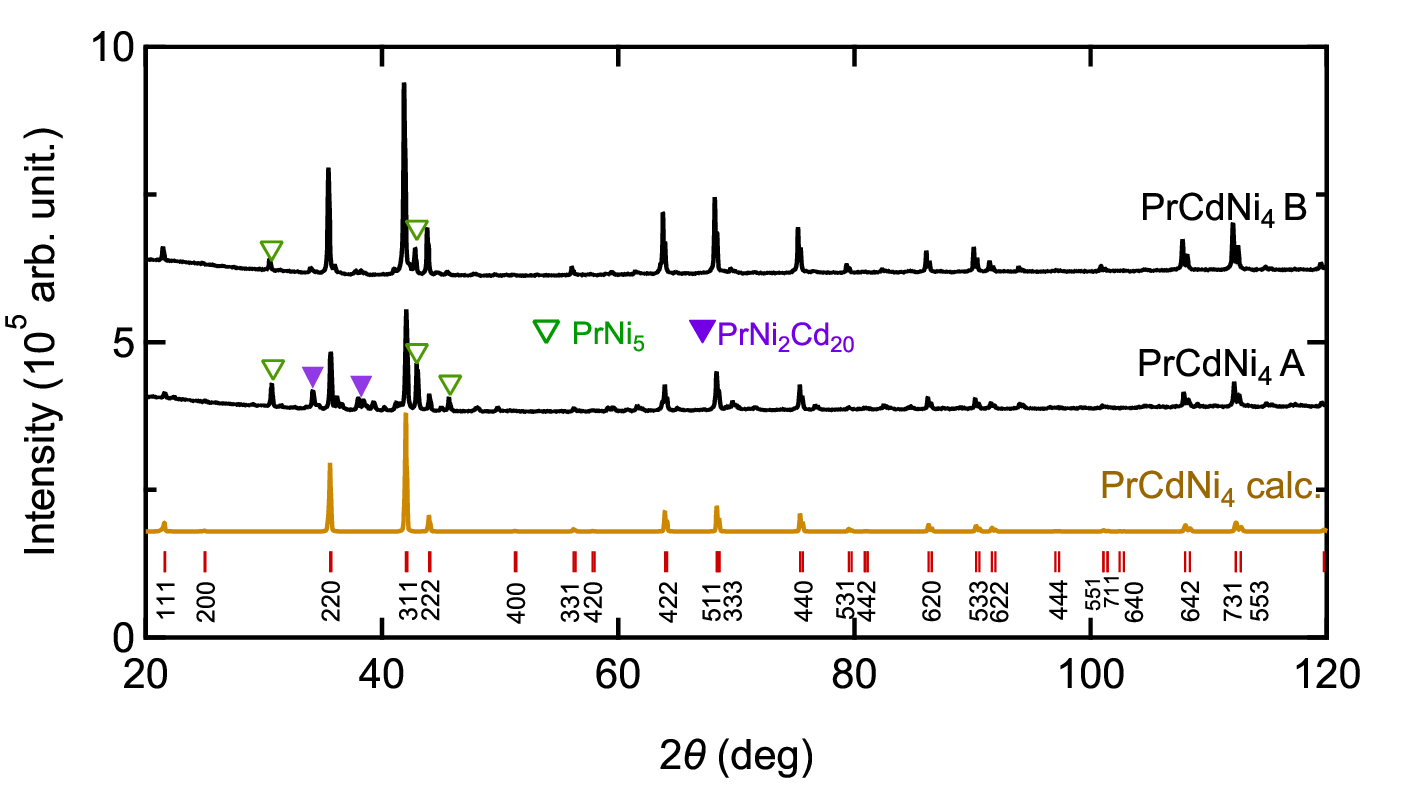}
\caption*{FIG. S2. 
Powder X-ray diffraction patterns of PrCdNi$_4$ (A) and (B).
The lower is the simulated pattern with the cubic MgSnCu$_4$-type structure.
Both samples can be indicated by cubic MgSnCu$_4$-type structure, whire additional peaks due to impurity phases shown with triangle.
}
\label{FigS02}
\end{figure}

The impurity phase of PrNi$_{5}$ was also observed in the powder neutron diffraction pattern at $T$ $=$ 2.0 K, as indicated by a star in Fig. S3.
We conducted the Rietveld analysis considering the two phases, PrCdNi$_{4}$ and PrNi$_{5}$, and determined their molar ratio to be PrCdNi$_{4}$:PrNi$_{5}$ $=$ 98.1(1) : 1.9(1).
It is noted that the peaks around 2$\theta$ $=$ 64 and 75 deg. were excluded from the analysis because they were due to aluminum from the sample container. 
Since neutrons can penetrate the samples, this result provides a reasonable stoichiometric ratio for the bulk samples in the macroscopic measurements. 

\begin{figure}[H]
\centering
\includegraphics[width=25pc]{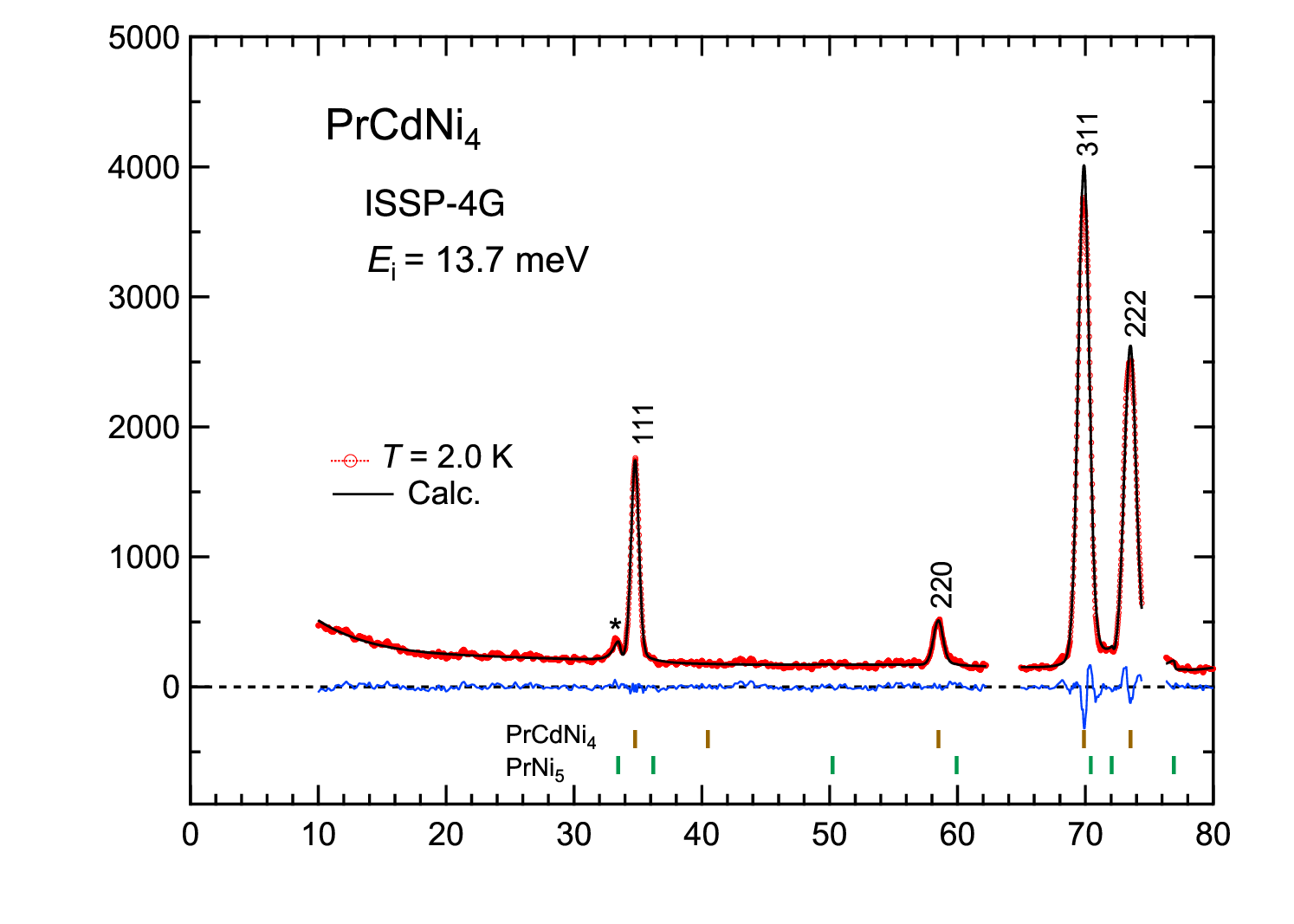}
\caption*{FIG. S3. 
Powder neutron diffraction patterns of PrCdNi$_4$ at $T$ $=$ 2.0 K.
The black line represents the fits to the data with the Rietveld analysis.
The blue line shows the difference between the data and calculation.
The brown and green bars indicate the scattering angles for the nuclear Bragg peaks of PrCdNi$_{4}$ and PrNi$_{5}$. 
}
\label{FigS03}
\end{figure}

\newpage

In Table S1, the structural parameters of PrCdNi$_4$ obtained from the Rietveld profile fitting analysis of neutron powder diffraction patterns for $T$ $=$ 2.0 and 0.32 K, shown with the black solid lines in Fig. 7 of the main text, are presented, along with the reliable factors for the fitting.
The lattice parameter at 0.32 K decreases by 0.23\% compared with the value at 273 K determined by the single-crystal X-ray analysis described in the next section.

\begin{table}[H]
\renewcommand{\arraystretch}{1.0}
\renewcommand{\baselinestretch}{1.2}%
\caption*{Table S1: Structural parameters of PrCdNi$_4$ refined by the Rietveld profile fitting analysis of neutron powder diffraction patterns. The space group is $F\bar{4}3m$ (\#216), where the Pr atoms occupy the 4$a$ site at (0, 0, 0), the Cd atoms the 4$c$ site at (1$/$4, 1$/$4, 1$/$4), and the Ni atoms 16$e$ site at ($x_{\rm Ni}$, $x_{\rm Ni}$, $x_{\rm Ni}$) with $Z$ $=$ 4.}
\label{TableS01}
\begin{center}
\small
\begin{tabular}{lll}
\hline
 & 0.32 K\hspace{12mm} & 2.0 K \\
\hline
Lattice parameters \hspace{7mm} & & \\
\hspace{3mm}$a$ (\AA) & 7.1112(3) &  7.1114(1) \\
\hspace{3mm}$V$ (\AA$^{3}$) & 359.61(3) & 359.644(9) \\
\hspace{3mm}$x_{\rm Ni}$ &  0.6278(4) & 0.625(2) \\
Reliable factor & & \\
\hspace{3mm}$R_{\rm p}$ & 4.74 & 5.20  \\
\hspace{3mm}$R_{\rm wp}$ & 6.24 &  6.75   \\
\hspace{3mm}$R_{\rm e}$ & 5.40 & 5.36  \\
\hspace{3mm}$S$ & 1.15 &  1.26 \\
\hline
\end{tabular}
\end{center}
\end{table}

\newpage

\section{Single-crystal X-ray structural analysis}

To identify the crystal structure, 
the single-crystal x-ray structural analysis was performed at 273 K with Mo $K{\alpha}$ radiation, $\lambda$ $=$ 0.071073 nm, monochromated by a multilayered confocal mirror using a Bruker APEX-II ULTRA CCD area-detector diffractometer at N-BARD, Hiroshima University. 
Crystallographic parameters and details for the measurement, data collection, and refinement of the single-crystal X-ray diffraction experiment are described in Tables S2 and S3.

\begin{table}[H]
\renewcommand{\arraystretch}{1.0}
\renewcommand{\baselinestretch}{1.0}%
\caption*{Table S2: Information of the measurement, data collection, and refinement of the single-crystal x-ray diffraction experiment for PrCdNi$_4$.}
\label{TableS2}
\begin{center}
\small
\begin{tabular}{lc}
\hline
Crystal system &  Cubic MgSnCu$_4$ \\
Space group & $F\bar{4}3m$ (\#216) \\
$a$ (\AA) & 7.12758(4)\\
$V$ (\AA$^3$) & 362.098(3) \\
Z & 4  \\
\\
Dimensions (mm$^3$)  & 0.21 $\times$ 0.14 $\times$ 0.08\\
Temperature (K) & 293 \\
Radiation, $\lambda$ (nm) & Mo-$K{\alpha}$, 0.071073 \\
$2\theta$ range ($^{\circ}$) & 9.908 -- 89.674 \\
$\mu$ (mm$^{-1}$) & 39.061 \\
\\
Data collection  & \\
Measured reflections  & 15017 \\
Unique reflections  & 186 \\
$h$ & $-$14 $\le$ $h$ $\le$ 14 \\
$k$ & $-$14 $\le$ $k$ $\le$ 14 \\
$l$ & $-$14 $\le$ $l$ $\le$ 14 \\
\\
Refinement  & \\
$\Delta_{\rho}$max$/$$\Delta_{\rho}$min (eA$^{-3}$) & 1.50$/$$-$1.62 \\
GoF & 1.257 \\
Reflections$/$parameters\hspace{5mm}  & 186/7 \\
$R_1$ & 0.0157 \\
$wR_2$ & 0.0412 \\
\hline
\end{tabular}
\end{center}
\end{table}

\begin{table}[H]
\renewcommand{\arraystretch}{1.0}
\renewcommand{\baselinestretch}{1.2}%
\caption*{Table S3: Crystallographic parameters for PrCdNi$_4$ determined at 273 K. $U_{\rm eq}$ is the isotropic displacement parameter defined as 1/3 of the trace of the orthogonalized $U_{ij}$ tensor.}
\label{TableS3}
\begin{center}
\small
\begin{tabular}{lllllll}
\hline
\multicolumn{5}{l} {Cubic MgSnCu$_4$-type} \\
\multicolumn{5}{l} {Space group: $F\bar{4}3m$ (\#216)} \\
\multicolumn{5}{l} {$a$ $=$  7.12758(4) \AA, $V$ $=$ 362.098(3) \AA$^3$, $Z$ $=$ 4} \\
\hline
Atom\hspace{3mm}  & Site\hspace{10mm} & $x$ & $y$ & $z$ & Occ.\hspace{7mm} & $U_{\rm eq}$ (\AA$^2$) \\ 
\hline
Pr & 4$a$ & 0 & 0 & 0 & 1 & 0.0071(2) \\ 
Cd & 4$c$  &  0.25 & 0.25  & 0.25 & 1 & 0.0041(2)  \\ 
Ni & 16$e$ & 0.62376(5)\hspace{3mm} & 0.62376(5)\hspace{3mm} & 0.62376(5)\hspace{3mm}  & 1 & 0.0052(2) \\ 
\hline
\end{tabular}
\end{center}
\end{table}

\newpage

\section{Phonon contribution to the specific heat}

Figure S3 shows the temperature dependence of the specific heat $C(T)$ of PrCdNi$_4$ from 2 K to 300 K.
%
%
The $C(T)$ data increase with elevating temperatures and approach the Dulong-Petit value of 3$nR$ = 149.7 J/K mol, where $n$ $=$ 6 is the number of atoms per formula unit and $R$ the gas constant. 
The inset displays the $C/T^3$ plots with the logarithmical scale.
The $C/T^3$ data increase on cooling. 
The (red) solid curve is a fit to the $C/T^3$ data between 50 and 300 K by adopting the Debye model for acoustic phonon modes \cite{Kittel05} with the Debye temperatures of $\theta_{\rm D}$ $=$ 270.7(6) K.
Since the acoustic phonon contribution to $C(T)$ is proportional to $T^3$ at lower temperatures, the calculated $C/T^3$ curve approaches a constant on cooling below 20 K.
Thereby, the difference between the data and the calculated curve for $T$ $<$ 50 K is ascribed to the electronic and magnetic contributions due to, respectively, the conduction electrons and the 4$f^2$ electrons of the Pr ion. 
The value of $\theta_{\rm D}$ $=$ 270.7(6) K is lower than that of $\theta_{\rm D}$ $=$ 300.3(6) K for an isostructural LaMgNi$_4$ \cite{Kusanose19}. 
This is reasonable because the molecular weight of PrCdNi$_4$ is much heavier than that of LaMgNi$_4$.

\begin{figure}[hb]
\centering
\includegraphics[width=20pc]{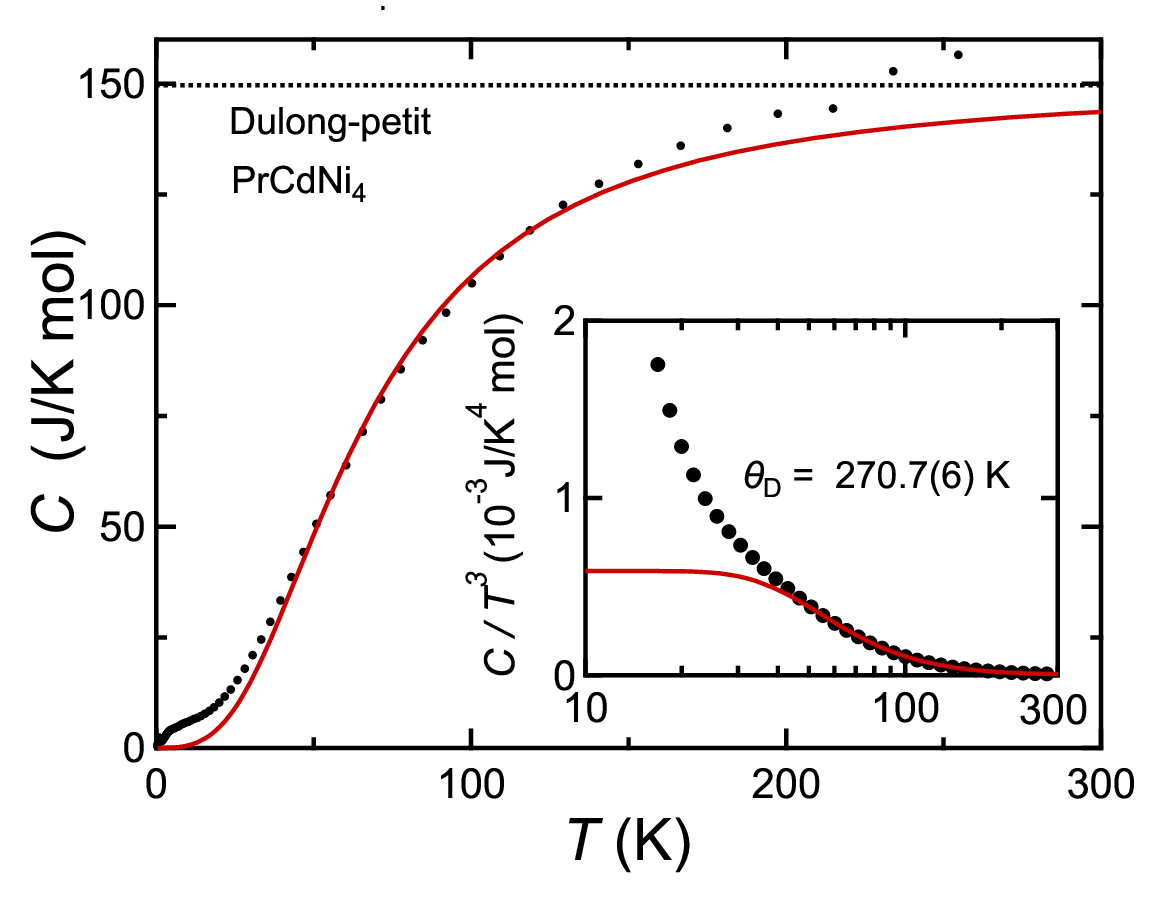}
\caption*{FIG. S4. Temperature variation of the specific heat of PrCdNi$_4$. The inset shows the $C/T^3$ plot with respect to the logarithmic temperature scale. 
The (red) solid lines are simulated by adopting the Debye model with a Debye temperature of $\theta_{\rm D}$ = 270.7(6)  K.
}
\label{FigS04}
\end{figure}